\documentclass[reprint,amsmath,amssymb,aps,prl,lengthcheck,floatfix,twocolumn,showpacs]{revtex4-2}

\usepackage{graphicx}
\usepackage{dcolumn}
\usepackage{bm}
\usepackage{easyReview}

\usepackage[normalem]{ulem} 
\newcommand\redout{\bgroup\markoverwith
{\textcolor{red}{\rule[.5ex]{2pt}{0.4pt}}}\ULon}
\usepackage{xcolor}

\begin{document}

\title{Absence of off-diagonal long-range order in hcp $^{\bf 4}$He dislocation cores}

\author{Maurice de Koning}
 \affiliation{Instituto de F\'{i}sica Gleb Wataghin, Universidade Estadual de Campinas, UNICAMP, 13083-859, Campinas, S\~{a}o Paulo, Brazil}
 \affiliation{Center for Computing in Engineering \& Sciences, Universidade Estadual de Campinas, UNICAMP, 13083-861, Campinas, S\~{a}o Paulo, Brazil}
 \email{dekoning@unicamp.br}
 
\author{Wei Cai}
\affiliation{Department of Mechanical Engineering, Stanford University, Stanford, CA 94305-4040}
\email{caiwei@stanford.edu}

\author{Claudio Cazorla}
\affiliation{Departament de Física, Universitat Politècnica de Catalunya, Campus Nord B4-B5, 08034 Barcelona, Spain}
\email{claudio.cazorla@upc.edu}

\author{Jordi Boronat}
\affiliation{Departament de Física, Universitat Politècnica de Catalunya, Campus Nord B4-B5, 08034 Barcelona, Spain}
\email{jordi.boronat@upc.edu}

\begin{abstract}
The mass transport properties along dislocation cores in hcp $^4$He are revisited by considering two types of edge dislocations as well as a screw dislocation, using a fully correlated quantum simulation approach. Specifically, we employ the zero-temperature path-integral ground state (PIGS) method together with ergodic sampling of the permutation space to investigate the fundamental dislocation core structures and their off-diagonal long-range order properties. It is found that the Bose-Einstein condensate fraction of such defective $^4$He systems is practically null ($\le 10^{-6}$), just as in the bulk defect-free crystal. These results provide compelling evidence for the absence of intrinsic superfluidity in dislocation cores in hcp $^4$He and challenge the superfluid dislocation-network interpretation of the mass-flux-experiment observations, calling for further experimental investigation.

\end{abstract}
\maketitle

Although torsional oscillator experiments on hcp $^4$He by Kim and Chan in 2004~\cite{Kim2004,Kim2004a} initially pointed at the existence of superfluidity in a solid-phase system, also known as supersolidity~\cite{Balibar2007,Boninsegni2012a}, posterior examination unambiguously established that, instead, the observed phenomenology was a consequence of its anomalous mechanical behavior. Specifically, it was found to be caused by the obstructing influence of $^3$He impurities on the low-temperature mobility of lattice dislocations~\cite{Kim2012,Day2007,Reppy2010,Haziot2013,Chan2013,Beamish2020}, the one-dimensional defects whose motion induces plastic deformation in crystalline solids~\cite{Hirth1992,Hull2001}.  

Still, the possibility of intrinsic supersolidity in hcp $^4$He has not been discarded, 
in particular due to a variety of mass flux experiments that report the flow of matter across solid $^4$He samples ~\cite{Ray2008,Ray2009,Ray2011,Vekhov2014,Vekhov2014a,Cheng2015,Cheng2016,Shin2017,Shin2019,Hallock2019}. 
However, the interpretation of these observations remains controversial. On the one hand, it has been proposed that the matter flow is transmitted through a superfluid network of interconnected, one-dimensional dislocation cores~\cite{Shin2017,Shin2019,Hallock2019}.  This view relies fundamentally on the results of computational grand-canonical finite-temperature path-integral Monte Carlo (PIMC) studies of one group~\cite{Boninsegni2007,Soyler2009,Kuklov2022}, which conclude that the cores of dislocations with Burgers vectors along the $c$-axis, $\textbf{b}=[0001]$, are superfluid at ultralow temperatures of $\sim 0.1$~K. In contrast, other authors argue that the mass flow is not dislocation-based but, rather, involves interfacial disorder effects within the samples, including at cell walls and grain boundaries~\cite{Cheng2015,Cheng2016}. This account is supported by the fact that large amounts of $^3$He impurities, much larger than required to saturate typical dislocation networks and their intersections, are required to block the flow at low temperatures~\cite{Cheng2016}.  In either case, dislocations play a central role in this controversy and, in view of the scarce computational evidence, further theoretical scrutiny of their properties is pressingly needed. 

In this Letter we  do so, revisiting the basic properties of dislocations in hcp $^4$He using first-principles quantum simulations.  However, the employed computational approach differs significantly from that applied in Refs.~\cite{Boninsegni2007}, \cite{Soyler2009} and ~\cite{Kuklov2022}. First, instead of finite-temperature PIMC calculations, we resort to the zero-temperature path-integral ground state (PIGS) approach, a generalization of the PIMC method to zero temperature~\cite{Sarsa2000,Rossi2009,Cazorla2017}, that has shown to converge to exact ground-state results regardless of the initially chosen wave function for condensed phases of $^4$He~\cite{Rossi2009,Rota2010}. Like in Refs.~\cite{Boninsegni2007,Soyler2009,Kuklov2022}, permutation sampling is carried out using the worm algorithm~\cite{Boninsegni2006b,Boninsegni2006c} to guarantee ergodicity in permutation space~\cite{Cazorla2017}. Second, we adopt different boundary conditions for the computational cells~\cite{Bulatov2006}. The results of the previous PIMC calculations~\cite{Boninsegni2007,Soyler2009} are based on tube-like setups, in which only atoms within a cylindrical (or pencil-shaped in the case of Ref.~\cite{Boninsegni2007}) region are treated explicitly while fixing a set of atoms outside of it to their classical positions, applying periodic boundary conditions (PBC) only along the dislocation line. Such an arrangement can give rise to lateral incompatibility stresses~\cite{Gehlen1972,Hoagland1976,Sinclair1978,Rao1998,Woodward2002} that may result in incorrect dislocation core structures if these are not adequately relieved, e.g., by using Green’s function boundary conditions~\cite{Rao1998,Woodward2002}. 
Here, we employ different configurations, including a dislocation-dipole arrangement employing fully three-dimensional PBC~\cite{Bulatov2006,LandinezBorda2016}, as well as slab configurations containing a single dislocation subject to two-dimensional PBC~\cite{Freitas2018}. Finally, we focus on the fundamental atomic lattice structure of the dislocation cores, without considering processes that require the addition or removal of material through a grand-canonical (GC) approach as used in Refs.~\cite{Boninsegni2007}, \cite{Soyler2009} and ~\cite{Kuklov2022}. Indeed, if one does not adequately thermalize, changing particle numbers may introduce artificial disorder, possibly leading to spurious the appearance of long-winding permutation cycles~\cite{Boninsegni2007}. By applying this computational scheme to edge dislocations with their Burgers vectors both perpendicular and parallel to the $c$ axis and to the screw dislocation with its Burgers vector along the $c$-axis, we find that, at zero temperature, the off-diagonal long-range order (ODLRO) is practically null ($\le 10^{-6}$), just as in the defect-free hcp crystal. This result contrasts previous claims~\cite{Boninsegni2007,Soyler2009,Kuklov2022} and signals the absence of quantum mass transport through dislocation cores in hcp $^4$He, instead lending support to the interpretation that the mass-flow observations are due to interfacial disorder effects rather than dislocation-mediated superfluidity. 

The integral Schr\"odinger equation for a system of $N$ interacting particles can be expressed in imaginary time as
\begin{equation}
\Psi({\bm R},\tau)=\int d {\bm R}^\prime \ G({\bm R},{\bm R}^\prime; \tau)
\Psi({\bm R}^\prime,0)~,
\label{green1}
\end{equation}
where $G({\bm R},{\bm R}^\prime; \tau) \equiv \langle {\bm R} | e^{-H \tau} | {\bm R}^\prime \rangle$ is the corresponding Green's function, with $H$ the system Hamiltonian, $\Psi({\bm R},\tau)$ the system wave function at imaginary time $\tau$ and $| {\bm R} \rangle = | {\bm r}_1, {\bm r}_{2}, \ldots, {\bm r}_N \rangle$, with ${\bm r}_i$ the particle positions. In the path-integral ground state (PIGS) approach~\cite{Sarsa2000,Rossi2009,Cazorla2017}, one exploits the formal identity between $G({\bm R},{\bm R}^\prime; \tau)$ and the thermal density matrix of the system at an inverse temperature of $\epsilon \equiv 1/T$ (we measure energy in units of Kelvins according to 1 K = 8.617$\times10^{-5}$ eV, such that $\hbar^2/2m$\,=\,6.059615 K\AA$^2$), namely, $\rho({\bm R},{\bm R}^\prime; \epsilon)$. In this manner, the ground-state wave function of the system, $\Psi_{0} ({\bm R})$, can be asymptotically projected out of a trial wave function, $\Psi_{T}({\bm R})$, according to
\begin{equation}
\Psi_{0} ({\bm R}_{M}) = \int \prod_{i=0}^{M-1} d{\bm R}_{i} ~\rho({\bm R}_{i},{\bm R}_{i+1};\epsilon) ~\Psi_{T} ({\bm R}_{0})~. 
\label{green2}
\end{equation}
Likewise, the ground-state average value of any physical observable can be written in terms of a multidimensional integral that can be calculated \textit{exactly},  within  statistical uncertainties, independently of whether the corresponding operator commutes or not with the Hamiltonian of the system. The only requirement for the trial wave function $\Psi_{T}$ is to satisfy the symmetry conditions imposed by the statistics of the simulated quantum many-body system. In this work, since we are dealing with boson particles, we consider a symmetrized trial wave function of the Jastrow type that typically is employed in quantum Monte Carlo (QMC) simulation of quantum liquids~\cite{Cazorla2017}.

The central physical quantity in our PIGS study is the one-body density matrix (OBDM), which is defined as
\begin{equation}
\rho_{1} ({\bm r}_{1}, {\bm r}^{\prime}_1) = \frac{1}{Z} \int d{\bm r}_{2} \ldots d{\bm r}_{N}   ~\rho({\bm R},{\bm R}^\prime)~,   
\label{onebody}    
\end{equation}
where the two configurations $|{\bm R} \rangle = |{\bm r}_1, {\bm r}_{2}, \ldots, {\bm r}_N \rangle$ and $|{\bm R}^\prime \rangle = |{\bm r}^{\prime}_1, {\bm r}_{2}, \ldots, {\bm r}_N \rangle$ differ only in one particle coordinate, and $Z$ represents the quantum partition function of the system. In PIGS, $\rho_{1} ({\bm r}_{1}, {\bm r}^{\prime}_1)$ is computed  by tracking the distances between the two extremities of one open chain (worm) during the QMC sampling~\cite{Ceperley1995}. Importantly, the condensate fraction of a $N$-boson system, $n_{0}$, can be deduced from the long-range asymptotic behavior of the OBDM,
\begin{equation}
    n_{0} = \lim_{|{\bm r}_1 - {\bm r}^{\prime}_{1}| \to \infty} \rho_{1} ({\bm r}_{1}, {\bm r}^{\prime}_1)~.
\label{condensate}
\end{equation}

\begin{figure}[t]
	\centering
	\includegraphics[width=1.00\linewidth]{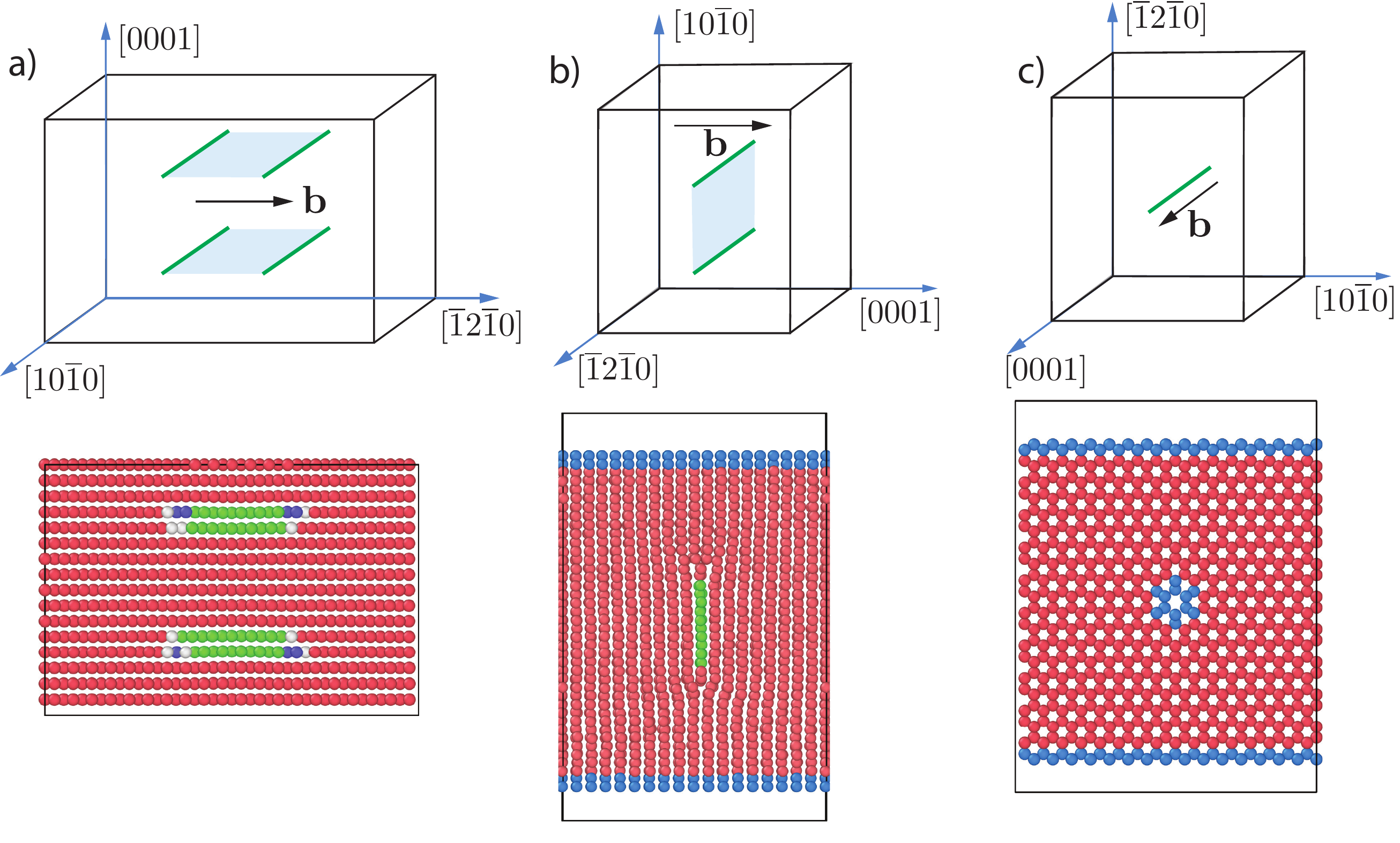}
	\caption{\label{Fig1} Computational cells employed in the zero-temperature PIGS simulations of edge and screw dislocations in hcp $^4$He as visualized using the OVITO package~\cite{Stukowski2010}. Atoms shown in red and green are located in hcp and fcc surroundings, respectively. In  all panels, the total Burgers vector $\bf{b}$ is indicated by the black arrow  a) Dipole arrangement for the  edge dislocations with Burgers vector in the basal plane, with PBC applied in all directions, following Ref.~\cite{LandinezBorda2016}. Each dislocation is dissociated into Shockley partial dislocations separated by a ribbon of stacking fault. b) Setup for the single edge dislocation with Burgers vector oriented along the $c$-axis dissociated into two Frank partials, with PBC applied along the dislocation line as well as the $c$-axis. The blue spheres in the upper and lower regions depict frozen atoms. c) Setup for single screw dislocation with Burgers vector oriented along the $c$-axis, with PBC applied along the dislocation line as well as the $[10\overline{1}0]$ directions. The blue spheres in the upper and lower regions depict frozen atoms. Blue atoms in the central region are close to the dislocation core.}
\end{figure}

We carried out PIGS simulations of hcp $^{4}$He crystals containing edge dislocations with their lines in the basal plane and with Burgers vectors oriented in the basal plane and along the $c$-axis, respectively, as well as for the screw dislocation with Burgers vector parallel to the $c$ axis. The interactions between He atoms were modeled using  the pairwise Aziz potential~\cite{Aziz1987}. The computational cells employed in the calculations are shown in Fig.~\ref{Fig1}. Depending on the type of edge dislocation, two different setups were employed. Fig.~\ref{Fig1}~a) displays the arrangement utilized for the basal edge (BE) dislocation. It is analogous to that used in Ref.~\cite{LandinezBorda2016}, containing a pair of edge dislocations  with opposite Burgers vectors of the type $\mathbf{b}=\tfrac{1}{3}[1\overline{2}10]$ dissociated into Shockley partials~\cite{Hirth1992} with Burgers vectors of the kind $\mathbf{b}=\tfrac{1}{3}[1\overline{1}00]$ separated by a stacking-fault ribbon. PBC were applied in all three directions and the cell contained 1872 atoms. As shown in Fig.~\ref{Fig1}~b), a different approach was adopted for the $c$-axis edge (CE) dislocation with Burgers vector $\mathbf{b}=[0001]$. While a dipole setup would also be possible, it would require simulating numbers of atoms that are prohibitively large for the excessively demanding PIGS calculations. Therefore, we employed a cell containing only a single CE dislocation, applying PBC along the dislocation-line direction and the $c$-axis while fixing the top and bottom two layers in the $[10\overline{1}0]$ directions. This is a standard approach that has been routinely used in atomistic simulations of dislocations~\cite{Freitas2018,AbuOdeh2022,Rodney2000} and preserves translational symmetry along the glide direction. The cell contains a total of 2280 atoms, of which 2052 were treated explicitly, whereas the remaining 228 atoms were fixed in the top and bottom layers. The CE dislocation dissociates into two Frank partial dislocations with Burgers vectors of the type $\mathbf{b}=\tfrac{1}{6}[20\overline{2}3]$ (Ref.~\cite{Hirth1992}, pg. 361) separated by a ribbon of stacking fault.  A similar single-dislocation setup was also employed for the $c$-axis screw (CS) dislocation, as shown in Fig.~\ref{Fig1}~c), with a cell containing 1920 of which 228 atoms in the surface layers were held fixed. For all dislocation cells the atomic number density was held fixed at  $\rho = 0.0287$~\AA$^{-3}$, which corresponds to a lattice parameter of $a=3.67$\AA . The  number of time-slices used in Eq. (2) was $M=25$ and an imaginary-time step of $\tau=0.0125$ K$^{-1}$. We have verified that larger values of $M$ and smaller values of  $\tau$do not modify our results within the statistical uncertainties (see Supporting Information~\cite{SupplementalMaterial}). Finally, for comparison with the defect-cell results, we also carried out subsidiary calculations for defect-free hcp $^4$He at the same density, employing a fully periodic cell containing 180 atoms.

\begin{figure}[t]
	\centering
	\includegraphics[width=1.00\linewidth]{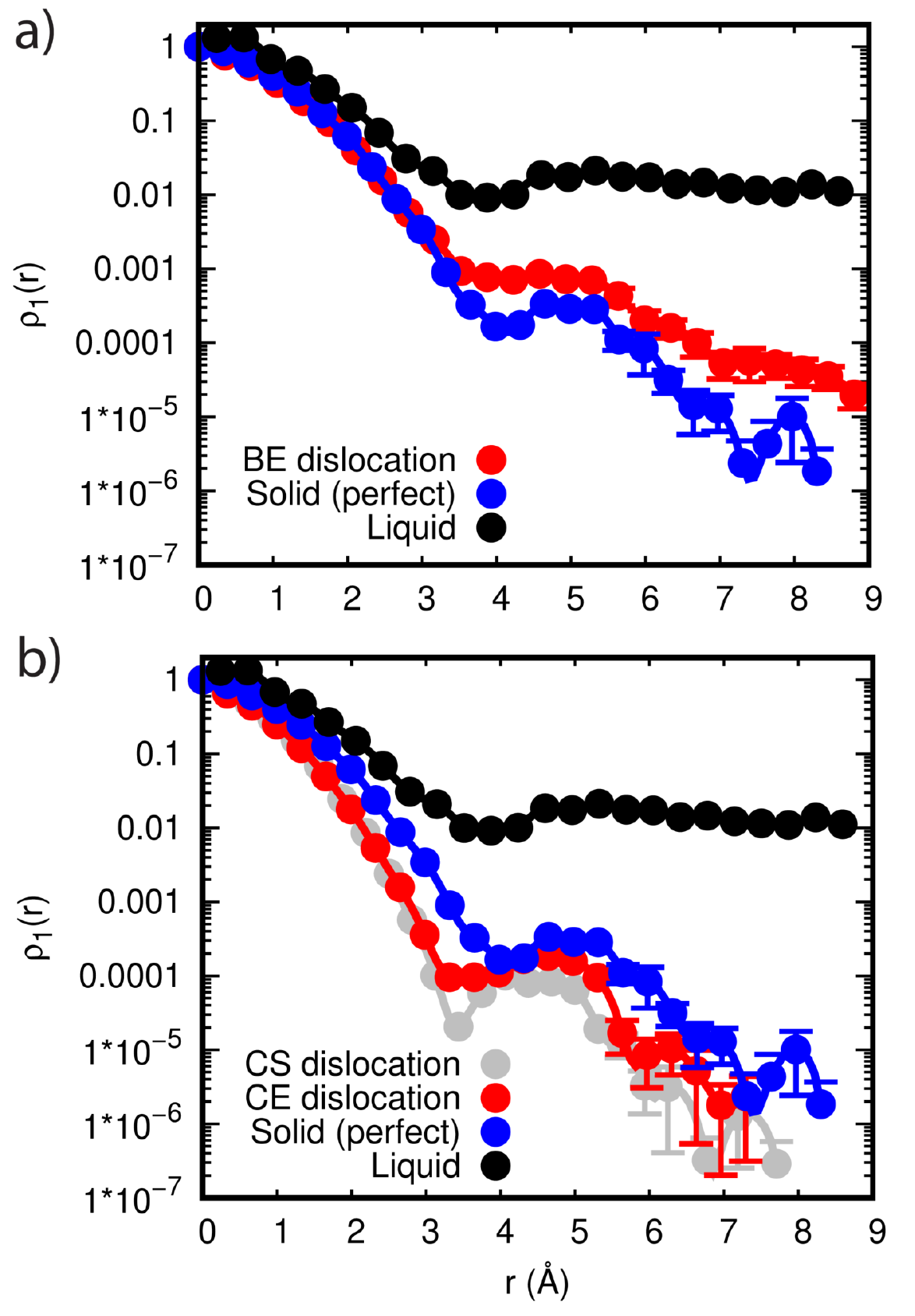}
	\caption{\label{Fig2} PIGS one-body density matrix results obtained at zero temperature for hcp $^4$He for the cells containing (a) a BE dislocation (red circles) and (b) CS (grey circles)  and CE dislocation (red circles). The $y$-axis is in logarithmic scale. For comparison, PIGS results obtained for defect-free bulk hcp $^4$He at the same density (blue circles) as well as the liquid at a density of  $0.0227$~\AA$^{-3}$ (black circles) are also shown.}
\end{figure}

The red circles in Figs.~\ref{Fig2}~a) and the red and grey circles in Fig.~\ref{Fig2}~b) show the PIGS results for the zero-temperature OBDM of hcp $^{4}$He crystals containing, respectively, the BE, CS and CE dislocations. In  all cases, $\rho_{1}$ clearly exhibits a  generally decreasing tendency under increasing radial distance $r \equiv |{\bm r}_1 - {\bm r}^{\prime}_{1}|$ (note the logarithmic $y$-scale in the graphs). For the BE dislocation, the steady OBDM reduction is slightly smaller than for the CS and CE dislocations; for example, at a radial distance of $\sim 7$~\AA~ the one-body density matrix has reduced to $\sim 10^{-5}$ in the former case compared to $\sim 10^{-6}$ for the latter. Nevertheless, the slope of  all $\rho_{1}$ asymptotes are manifestly negative. This is clear evidence that the Bose-Einstein condensate fraction (Eq.\ref{condensate}) of bulk hcp $^{4}$He containing  these types of dislocations is negligible in practice ($\le 10^{-6}$) as $\rho_{1}$ tends to zero in the limit of long radial distances. For further comparison, the blue circles in Figs.~\ref{Fig2} a) and b) display  the PIGS OBDM calculations carried out for the defect-free hcp $^4$He cell at the same density. 

 The results for these dislocation systems display the same general trend as seen for the defect-free crystal, providing further support for our conclusion of negligible $n_{0}$ in the presence of these types of  dislocations. As a final consistency check, we carried out an additional simulation starting from the CE dislocation cell, but reducing its density to $0.0227$~\AA$^{-3}$ to induce a transition into the liquid phase. The corresponding ODLRO, obtained after reaching the equilibrated liquid, is shown as the black circles in Fig.~\ref{Fig2}~b). The Bose-Einstein condensate fraction obtained in this case, employing the same PIGS approach applied to the solid-phase systems, is found to be $n_{0} \sim 0.02$.  This is in agreement with the known value corresponding to bulk liquid $^4$He at that density at ultralow temperatures~\cite{Rota12}, attesting to the numerical reliability of our zero-temperature computational approach.

\begin{figure*}[t]
	\centering
	\includegraphics[width=0.7\linewidth]{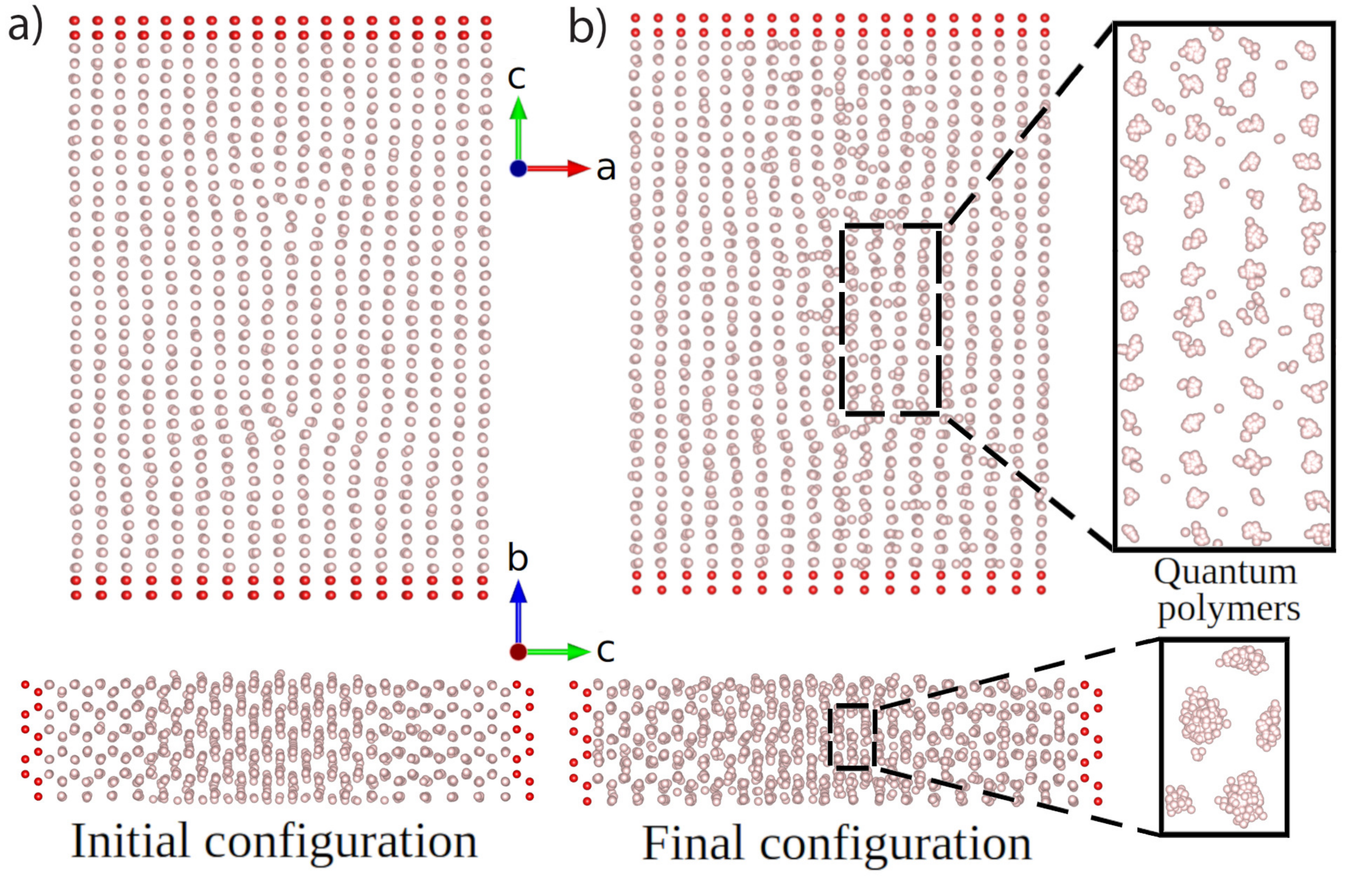}
	\caption{\label{Fig3} Visualization of the $^{4}$He system containing the dissociated CE dislocation at the beginning and end of the PIGS simulations; quantum polymers ``centroids'' are represented in both cases. The initial configuration was obtained after equilibrating the system at $T = 1$~K with the PIMC method. A few quantum polymers located at a similar distance within the dislocation core are represented in the inset of b); long chains of atomic exchanges involving several quantum polymers are absent.}
\end{figure*}

The fact that the zero-temperature OBDM results in Fig.~\ref{Fig2} display a practically null Bose-Einstein condensate fraction (i.e., $\lesssim 10^{-6}$) in both the defect-free as well as defected $^{4}$He crystal is compelling evidence that the cores of  the considered types of dislocations are in fact insulating in nature.  The lack of quantum mass flux along the  dislocation cores can be further verified by visual inspection of the quantum polymers during the simulation. A representative example is depicted in Fig.~\ref{Fig3} for the case of the dissociated CE dislocation. Fig.~\ref{Fig3}~a) and the main panel of Fig.~\ref{Fig3}~b)  display the  centroids (i.e., the ``centers-of-mass'' of the quantum polymers) for the initial and final configurations of the PIGS simulation, respectively. Both pictures qualitatively demonstrate the prevalence of atomic order, including the regions of the partial dislocation cores. Furthermore, when visualizing entire quantum polymers in the core region as depicted in the expanded view,  there are no evident traces of long-winding quantum exchanges~\cite{Ceperley1995},  thus corroborating the absence of superfluidity in these dislocation cores. 

While the absence of superfluidity for the BE dislocations is consistent with the PIMC calculations reported in Ref.~\cite{LandinezBorda2016} and the unpublished data referred to in Ref.~\cite{Pollet2008}, the present PIGS results for the CS and CE dislocations are at odds with the findings in Refs.~\cite{Boninsegni2007} and ~\cite{Soyler2009} as well as the proposed mechanism of  ``superclimb'' of dislocations~\cite{Soyler2009,Kuklov2022}. Accordingly, our results are incompatible with the superfluid dislocation network interpretation of the mass flux experiments, and lend support to the alternate view that effects related to disordered regions at internal interfaces, including vessel walls and grain boundaries, are responsible for the observations~\cite{Cheng2015,Cheng2016}.  

A further issue with the superfluid-network interpretation is that, given the consensus that dislocations with Burgers vectors in the basal plane are insulating~\cite{LandinezBorda2016,Pollet2008}, it relies fundamentally on the presence of a spanning network consisting entirely of dislocations with $c$-axis Burgers vectors. Such an arrangement of dislocations, however, is geometrically impossible due to the requirement of conservation of Burgers vector at network nodes~\cite{Hirth1992}. In contrast, there is abundant experimental evidence~\cite{Hiki1976,Hiki1977,Tsuruoka1979,Paalanen1981,Day2007,Haziot2013} for the existence of networks of nonsuperfluid basal-plane Burgers-vector dislocations, which drive the dominant mode of basal slip in hcp $^4$He~\cite{Tsuruoka1979,Paalanen1981} and play a central role in the phenomenon of giant plasticity~\cite{Haziot2013}, as well as in the nonsupersolid explanation of the original torsion-oscillator observations by Kim and Chan~\cite{Day2007}. This premise is also consistent with findings in other hcp-structured materials  such as Zn~\cite{Tyapunina1975} and Mg~\cite{Hirsch1965} in which observed dislocation networks display the characteristic hexagonal structure of basal-plane Burgers vector dislocations. In this light, the present results further challenge the superfluid dislocation-network interpretation of the mass-flux-experiment observations and call for further experimental investigation.

M.K. acknowledges support from CNPq, Fapesp grant no. 2016/23891-6 and the Center for Computing in Engineering \& Sciences - Fapesp/Cepid no. 2013/08293-7. W.C. acknowledges support from the U.S. Department of Energy, Office of Basic Energy Sciences, Division of Materials Sciences and Engineering under Award No. DE-SC0010412. J.B.  acknowledges financial support from the Secretaria d'Universitats i Recerca del Departament d'Empresa i Coneixement de la Generalitat de Catalunya, co-funded by the European Union Regional Development Fund within the ERDF Operational Program of Catalunya (project QuantumCat, Ref.~001-P-001644), and the MINECO (Spain) Grant PID2020-113565GB-C21. C.C. acknowledges financial support from the MINECO (Spain) under the ``Ram\'on y Cajal'' fellowship (RYC2018-024947-I).

\bibliographystyle{apsrev4-2}

%

\end{document}